\documentclass[aps,prl,twocolumn,groupedaddress,showpacs]{revtex4}
\usepackage{graphicx}
\usepackage{dcolumn}
\begin{document}
\title{Correlated Quantum Transport of Density Wave Electrons}

\author{J.~H.~Miller,~Jr.}
\email[]{jhmiller@uh.edu}
\affiliation{Department of Physics, University of Houston, Houston, Texas 77204-5005 USA}
\affiliation{Texas Center for Superconductivity, University of Houston, Houston, Texas 77204-5002 USA}
\author{A.~I.~Wijesinghe}
\affiliation{Department of Physics, University of Houston, Houston, Texas 77204-5005 USA}
\affiliation{Texas Center for Superconductivity, University of Houston, Houston, Texas 77204-5002 USA}
\author{Z.~Tang}
\affiliation{Department of Chemistry, University of Houston, Houston, Texas 77204-5003 USA}
\author{A.~M.~Guloy}
\affiliation{Texas Center for Superconductivity, University of Houston, Houston, Texas 77204-5002 USA}
\affiliation{Department of Chemistry, University of Houston, Houston, Texas 77204-5003 USA}

\date{\today}

\begin{abstract}
Recently observed Aharonov-Bohm quantum interference of period $h/2e$ in charge density wave rings strongly suggest that correlated density wave electron transport is a cooperative quantum phenomenon. The picture discussed here posits that quantum solitons nucleate and transport current above a Coulomb blockade threshold field.  We propose a field-dependent tunneling matrix element and use the Schr\"{o}dinger equation, viewed as an \textit{emergent classical equation} as in Feynman's treatment of Josephson tunneling, to compute the evolving macrostate amplitudes, finding excellent quantitative agreement with voltage oscillations and current-voltage characteristics in NbSe$_3$. A proposed phase diagram shows the conditions favoring soliton nucleation versus classical depinning.
\end{abstract}


\maketitle

Cooperative quantum tunneling has emerged as an important class of phenomena, whose manifestations include Josephson tunneling~\cite{Josephson1962251}, macroscopic quantum tunneling, and decay of the false vacuum~\cite{PhysRevD.15.2929}. The latter describes instability of a scalar field~$\phi(r)$, where ``vacuum" refers to a minimum energy state. In this Letter, $\phi$ represents the phase of a density wave. If~$\phi(r)$ sits in a metastable well (`false vacuum') it is unstable to decay by tunneling into a lower potential well within a small region, nucleating a bubble of `true vacuum' bounded by solitons~\cite{PhysRevLett.39.46}. Herein we propose coherent Josephson-like tunneling of $microscopic$ quantum solitons (single-chain solitons, delocalized in both longitudinal and transverse directions) of charge $\pm2e$ within a quantum fluid, i.e.~\textbf{$not$} macroscopic quantum tunneling (MQT) of a massive object.

The charge density wave~(CDW) exhibits a charge modulation $\rho\left(x, t\right)=\rho_0(x,t) + \rho_1 \cos[2k_Fx-\phi(x, t)]$ along the chain direction, while the spin density wave~(SDW) is equivalent to two out-of-phase CDWs for the spin-up and -down subbands~\cite{dwinsGG}. Like a superconductor, the density wave~(DW) is a correlated electron (or electron-phonon) system capable of collective charge transport. The superconducting condensate is a charged superfluid represented by a complex order parameter. Its behavior can be described by the Schr\"{o}dinger equation as an emergent classical equation for the condensate (Ch.~21 of~\cite{1965flp}). Unlike a superconductor, however, the order parameter corresponding to the DW charge or spin modulation does not couple directly to an electric field or vector potential. Nevertheless, \textit{gradients} or kinks in DW phase carry charge that: 1)~couple to an externally applied electric field, and 2)~generate their own electric field that leads to a Coulomb blockade effect. 

We propose that nucleated droplets of many $\pm2e$ charged kinks and antikinks behave as quantum fluids due to interchain interactions and quantum delocalization. We use the time-dependent  Schr\"{o}dinger equation to describe the coupled macrostates in a manner, as in Josephson tunneling, that is \textit{classically robust} against decoherence below the transition temperature. In the proposed Josephson-like tunneling process, the quantum fluid flows over or through the barrier over a long time scale (up to $\sim1\mu$s). Thermal excitations are frozen out by the Peierls gap since the condensate has one thermal degree of freedom within a phase-coherent domain~\cite{Bardeen1989}.

Aharonov-Bohm oscillations of period $h/2e$ in the CDW magneto-conductance of NbSe$_3$ crystals with columnar defects~\cite{PhysRevLett.78.919} and TaS$_3$ rings~\cite{Tsubota2009416} suggest cooperative quantum behavior, in some cases over distances of 85~ $\mu$m and for $T>$~77~K. Moreover, the $h/2e$, rather than $h/2Ne$ period predicted~\cite{PhysRevB.42.7614} for $N$ parallel chains, supports the idea of coherent Josephson-like tunneling of microscopic entities of charge~$2e$ within a quantum fluid, rather than MQT of a massive object. 

\begin{figure} [ht] 
\includegraphics[scale=.5]{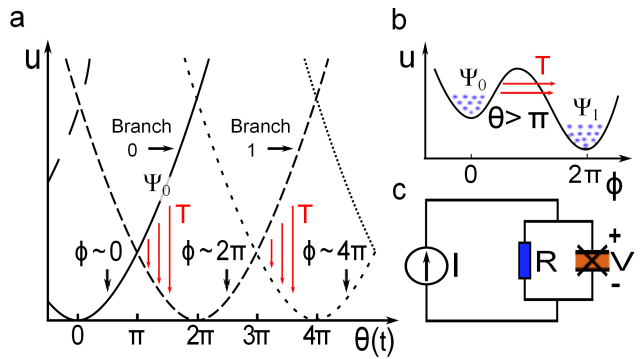}
\caption{(color online) \textbf{a}.~Potential energy vs. $\theta$  for $\phi \sim 2\pi n$.
\textbf{b}.~$u(\phi)$ when $\theta=2\pi E/E^* > \pi$  as the phases~$\phi_k(x)$ tunnel coherently into the next well. 
\textbf{c}.~Time-correlated soliton tunneling model.\label{fig1}}
\end{figure}

While the classical DW depinning field~$E_{cl}$ is well understood, less widely known is the existence of a Coulomb blockade threshold field~$E_T$ (smaller than~$E_{cl}$) above which the system becomes quantum mechanically unstable~\cite{1976AnPhy.101..239C}. This threshold is readily determined for nucleation of charge soliton pairs in 1-D~\cite{1976AnPhy.101..239C, Krive1985691, PhysRevLett.84.1555} or (in~3-D) soliton domain wall pairs, of charge $\pm Q_0=\pm2Ne\rho_c$  where~$\rho_c$ is the condensate fraction~\cite{0305-4470-36-35-308}. Just like the charged electrodes of a parallel-plate capacitor, these produce an internal field, $E^*=Q_0/\epsilon A$. If an external field~$E$ is applied, the difference in electrostatic energies with, $\frac{1}{2}\epsilon^2 (E\pm E^*)^2$, and without the pair, $\frac{1}{2}\epsilon E^2$, is positive when $\left|E\right| < E_T=\frac{1}{2}E^*$, yielding the Gr\"{u}ner relation~\cite{dwinsGG}: $\epsilon E_T \sim e\rho _cN/A$~\cite{comsol}. When added to the periodic pinning energy, this quadratic electrostatic energy ensures that the DW phase sits in the lowest potential energy well, or `true vacuum' state when $\left|E\right| < E_T$. However, when $E>E_T$, or  $\theta=2\pi E/E^*>\pi$, the formerly `true vacuum' becomes a metastable state or `false vacuum' (Figs.~\ref{fig1}a~\&~b).

Density waves usually have anisotropic relative dielectric responses, $\epsilon_{\|}~\gg~\epsilon_{\perp}$, vs. the chain direction. Using rescaled coordinates, $x'=x/\epsilon_{\|}$, etc., a single-chain dislocation pair looks like a parallel plate capacitor that produces a field $E^*=2e/2\epsilon A_{ch}$, where~$A_{ch}$ is the cross-sectional area of a DW chain and $\epsilon=\epsilon_{\|}\epsilon_0$~\cite{comsol}. Thus, the Coulomb blockade threshold is comparable to that for domain wall pair creation, within a factor of~$\sim 1/2$. Many nucleated $2\pi $ dislocations, of charge  $\pm2e$ each~\cite{2008SSSci..10.1786B}, can then form droplets with quantum fluidic properties. The temperature-dependence of $E_T$ goes inversely with that of $\epsilon (T)$~\cite{comsol}.

Our model relates the `vacuum angle'~$\theta$ (e.g. ~ref.~\cite{1976AnPhy.101..239C} and citing papers) to displacement charge~$Q$ between contacts by: $\theta=2\pi(Q/Q_0)$. The potential energy of the $k^{th}$ chain can then be written as~\cite{1976AnPhy.101..239C, PhysRevLett.84.1555}:	 
\begin{equation}
u\left[\phi_k\right] = 2u_0\left[1 - \cos \phi_k(x)\right] + u_{E}\left[\theta-\phi_k(x)\right]^2 
\end{equation}
where the first term is the periodic DW pinning energy. The quadratic term is the electrostatic energy resulting from the net displacement charge or, equivalently, the applied field and internal fields created by kinks due to phase displacements. Fig.~\ref{fig1}a plots $u$~vs.~$\theta$ when the energy is minimized for $\phi \sim 2\pi n$  (dropping the subscript) when~$u_E<<u_0$. The phases~ $\phi_k$ tunnel coherently into the next well via a matrix element $T$ (Fig.~\ref{fig1}b) as each parabola, or branch (Fig.~\ref{fig1}a), crosses the next at the instability points  $\theta=2\pi (n + 1/2)$. Here we propose an idealized time-correlated soliton tunneling model to simulate DW dynamics. It includes a shunt resistance~$R$, representing normal, uncondensed electrons, in parallel with a capacitive tunnel junction representing soliton tunneling (Fig.~\ref{fig1}c), by analogy to time-correlated singe-electron tunneling (SET)~\cite{BF00683469}. 

Advancing the phase of all parallel chains by~$2\pi n$ creates multiple pairs of soliton domain walls that quickly reach the contacts. Similar to SET, the voltage is then proportional to net displacement charge: $V=(Q-nQ_0)/C=(Q_0/2\pi C)[\theta-2\pi n]$  when the phase has advanced to $\left\langle \phi \right\rangle=2\pi n$ between the contacts. More generally, if the phase expectation value $\left\langle \phi\right\rangle$ among~$N$ parallel chains advances by a fraction or non-integer multiple of $2\pi$, the voltage is $V=(Q_0/2\pi C)[\theta-\left\langle \phi \right\rangle]$ where $C=\epsilon A/\ell$.  This leads to a total current: $I=I_n+I_{DW}$, where  $I_n=(Q_0/2\pi RC)[\theta-\left\langle \phi \right\rangle]$ is the normal current, and  $I_{DW}=\frac{dQ}{dt}=\frac{Q_0}{2\pi}\frac{d\theta}{dt}$ is the DW current. Defining $\omega=2\pi I/Q_0$  and $\tau\equiv RC$  yields the following equation for the time evolution of~$\theta$:
\begin{equation}
\frac{d\theta}{dt} =\omega-\frac{1}{\tau}[\theta-\left\langle \phi\right\rangle].
\end{equation}
We compute $\left\langle \phi\right\rangle$  by solving the Schr\"{o}dinger equation,
\begin{equation}
i\hbar \frac{\partial \psi_{0,1}}{\partial t}=U_0\psi_{0,1}+T\psi_{1,0}~, 
\end{equation}
to compute the original and emerging macrostate amplitudes~$\psi_0(t)$  and~$\psi_1(t)$  (more generally~$\psi_n$  and~$\psi_{n+1}$) for the system to be on branches ~$0$ and~$1$ (or~$n$ and~$n+1$, Fig.~\ref{fig1}a), respectively. We interpret these amplitudes to represent classically robust order parameters, and the above equation is viewed as an emergent classical equation following Feynman~\cite{1965flp}.  The macrostates are coupled via a tunneling matrix element~$T$ with a Zener-like field dependence.  (Another approach, employing probabilities rather than amplitudes, yields sharp sawtooth-shaped voltage oscillations and will be discussed elsewhere.)

Our model represents the amplitudes~$\psi_{0,1}$ by: $\psi_{0,1} =\sqrt{\rho_{0,1}}\exp\left[i\delta_{0,1}\right]$, where~$\rho_{0,1} =N_{0,1}/N$  is the fraction of parallel chains on the respective branch. Advancing~$\phi_k(x)$ by~$2\pi$  within a given region, taking~$\phi_k$ from one branch to the next, is equivalent to creating a pair of microscopic $2\pi$-solitons. Thus, the macrostate order parameters $\psi_{0,1}$ are coupled via coherent, Zener/Josephson tunneling of delocalized quantum solitons~\cite{PhysRevB.43.12205}, with an enormous aggregate of~$N$ (up to $\sim$ 10$^9$) such processes occurring coherently.

The driving force is the energy difference per unit length between potential minima at $\phi\sim2\pi n$  and $\phi\sim2\pi(n+1)$. When $\alpha \equiv u_E/u_0 << 1$, this force is given by: $F=4\pi u_E\theta_n'$, where~$\theta_n'=\theta- 2\pi(n+\frac{1}{2})$. Following Bardeen~\cite{PhysRevLett.6.57, Duke}, $T$ is estimated as:  $T(F)= -4F\lambda\exp[-F_0/F]$, where $\lambda^{-1} \sim \Delta_\varphi/\hbar v_0+\lambda_m^{-1} $, $\lambda_m$ is a mean free path length, $\Delta_\varphi$   is the microscopic soliton energy, $v_0$ is the phason velocity, and $F_0\sim \Delta_\varphi^2/\hbar v_0$. This expression is similar to the rate of Schwinger pair production in 1-D~\cite{PhysRevD.78.036008}. Since any negative energy difference (Figs.~\ref{fig1}a \& b) within the `bubble' along the $x$-~direction is balanced by the positive soliton pair energy at its boundaries, $T$ couples states of equal energy, $U_0=U_1=U$. Thus, defining $\psi_{0,1} =\chi_{0,1}(t)\exp[-iUt/\hbar]$, the Schr\"{o}dinger equation (Eq.(3)) reduces to: $i\hbar\partial{ \chi_{0,1}}/\partial t = T\chi_{0,1}$.

We define: $t'=t/\tau$, $f=\omega\tau/2\pi$  ($\propto I$), $q=\theta/2\pi, q_o=F_0/2F_T=\theta_0/2\pi, F_T=2eE_T$, and $q_n'=\theta_n'/2\pi=q-n-\frac{1}{2}$, to simplify the computations. Finally, setting  $\chi_0(t)=c_0(t)$ and $\chi_1(t)=ic_1(t)$, taking~$c_0$ and~$c_1$ to be real, yields the coupled equations:  
$dc_1/dt'=\left[\gamma q_n' \exp(-q_0/q_0')\right]c_0$    and   
$dc_o/dt'=-\left[\gamma q_n' \exp(-q_0/q_n')\right]c_1$   
for $q_n'>0$,  where $\gamma=32\pi^2u_E\lambda \tau/\hbar$.
These are integrated numerically, with initial values $c_0=1$ and $c_1=0$, yielding: $\left\langle \phi\right\rangle=2\pi[n+p]$, where  $p=\left|c_1\right|^2$.
The transition from branch $n$ to $n +1$ is considered complete, and $n$ is incremented while $p$ is reset back to zero, once~$p$ exceeds a cutoff close to one (e.g.~0.9995).
When an applied current pulse is turned off, any remaining displacement charge discharges back through the shunt resistance and the system retains a memory of the previous macrostate amplitudes.
The algorithm thus incorporates backward transitions from branch~$n$ to branch~$n-1$ when $dQ/dt$, $F$, and $q_n'$ are negative. We find that no more than three training pulses are needed to converge to a `fixed point' of one or two voltage oscillation patterns, which are averaged.

\begin{figure} [ht]
\includegraphics[scale=0.5]{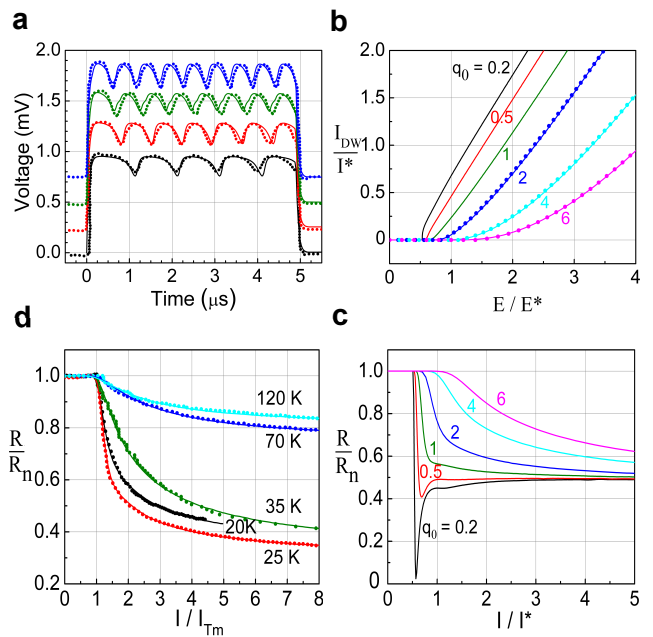}
\caption{(color online) \textbf{a}.~Theoretical (solid lines) vs. experimental (dashed lines~\cite{PhysRevB.61.10066}) voltage
 oscillations of an NbSe$_3$ crystal at 52~K for current pulse amplitudes (bottom to top, offset by~0, 0.25, 0.5, and~0.75~V for clarity): 9.90~$\mu$A, 10.89~$\mu$A, 11.49~$\mu$A, and 11.88~$\mu$A. \textbf{b}.~Simulated DW current vs. field for $\gamma$=~1.5 and several~$q_0$. Dotted lines (Table I):~Bardeen's modified Zener function~\cite{PhysRevLett.45.1978}. \textbf{c}.~Simulated $R=dV/dI$ vs. $I/I^*$,  where $I^*\equiv E^*\ell/R_n$ , where $R_n$ is the normal resistance at zero bias, for several $q_0$ and  $ \gamma$=~1.5. \textbf{d}.~Theoretical (solid lines, Table~\ref{table2}) vs. experimental (dotted lines) $dV/dI$ vs. $I$ for an NbSe$_3$ crystal.\label{fig2}}
\end{figure}

Fig.~\ref{fig2}a compares the quantum theory with measured voltage oscillations~\cite{PhysRevB.61.10066} of NbSe$_3$ for rectangular current pulses. The parameters used for the theoretical plots (solid lines) in Fig.~\ref{fig2}a are:  $\gamma$=0.5, $q_0$=0.7, $\tau$=51~ns, $V^*= E^*\ell$=1.11~mV (where~$\ell$=~distance between contacts), and~$R_n$=~99.6~$\Omega$. The measured threshold current of 6.93~$\mu$A is taken to correspond to $f$= 0.6, the normalized onset threshold current consistent with the chosen values of~$\gamma$  and~$q_0$, while the remaining normalized current pulse amplitudes~$f$ are scaled to the indicated amplitudes in Fig.~\ref{fig2}a.

The theoretical voltage oscillation amplitudes agree remarkably well with experiment, especially considering the simplicity of the model. Moreover, the model correctly reproduces the observed progression of non-sinusoidal shapes, ranging from rounded backward sawtooth behavior for the 9.90-$\mu$A current pulse to more symmetrical oscillations for higher current pulse amplitudes. The oscillation frequency is~$f=I_{DW}/Q_0$, so the number of oscillations per pulse, captured correctly by the model, increases with current. During each cycle, the CDW current, $I_{cdw} = I - V/R_n$, gradually increases over a significant portion of a cycle as the voltage decreases from its maximum value.  As seen in the bottom plot of Fig.~\ref{fig2}a, this time scale can be up to $\sim~1~\mu$s, supporting the idea that the quantum fluid flows through the barrier for a relatively long time.

The $I-V$ and $dV/dI$ curves are computed by averaging the voltage over several cycles, with results shown in Fig.~\ref{fig2}b \&~c. A range of behaviors are captured, with rounded Zener-like behavior (e.g.~Refs.~\cite{PhysRevLett.45.1978, PhysRevLett.55.1006}), emerging for large $q_0 \propto  E_0/E_T$, as contrasted with, when~$q_0$ is small, more linear $I-V$ curves and $dV/dI$ curves with negative dips or wings, as seen in NbSe$_3$ crystals with fewer impurities~\cite{ PhysRevLett.58.828}. For small~$q_0$, the `measured' threshold $E_{Tm}$ occurs near the Coulomb blockade threshold: $E_{Tm}=E_T=E^*/2$. However, $E_{Tm}$ becomes larger than $E^*/2$ as $q_0$ increases. The dotted lines in Fig.~\ref{fig2}b are obtained from a normalized Bardeen function~\cite{PhysRevLett.45.1978}, $I_{DW}/I^*=\Gamma[E'-E_{Tm}']\exp[-E_{0m}'/E']$, where~$E'=E/E^*$ and $I^*=E^*\ell /R_n$, while $E_{Tm}'=E_{Tm}/E^*$   and $E_{0m}'=E_{0m}/E^*$  are normalized `measured' threshold and Zener activation fields, and   $\Gamma=~1.0$ for all three plots. The remaining parameters used for the Bardeen function fits are shown in~Table~\ref{table1}.

\begin{table}[h]
\caption{\label{table1}Parameters used to generate the Bardeen function plots in Fig.~\ref{fig2}b.}
\begin{ruledtabular}
\begin{tabular}{lcr}
\textrm{$q_0$}    &
\textrm{$E_{Tm}'$}&
\textrm{$E_{0m}'$}\\
\colrule
2 &  0.847 & 0.96\\
4 &  1.10  & 2.55  \\
6 &  1.40  & 4.05\\
\end{tabular}
\end{ruledtabular}
\end{table}

The dotted lines in Fig.~\ref{fig2}d show, for an NbSe$_3$ crystal, differential resistance,~$R=dV/dI$, normalized to normal resistance $R_n=dV/dI|_{zero bias}$ vs. $I/I_{Tm}$, where~$I$ is total applied current and $I_{Tm}=V_{Tm}/R_n$ is the measured threshold current. The solid lines in Fig.~\ref{fig2}d are simulation results using the parameters indicated in Table~\ref{table2}. Fig.~\ref{fig2}d shows excellent agreement between theory and the~$dV/dI$ measurements, showing rounded behavior below the upper and lower Peierls transitions.

\begin{table}[h]
\caption{\label{table2}Parameters used to simulate the solid $dV/dI$ curves in Fig.~\ref{fig2}d.}
\begin{ruledtabular}
\begin{tabular}{lcdr}
\textrm{Temperature}   &
\textrm{$I_{Tm}/I^*$}  &
\textrm{$\gamma$}      &
\textrm{$q_0$}         \\
\colrule
~20~K   &  0.87 & 2.7   &  3.0 \\
~25~K   &  0.76 & 3.2   &  2.3 \\
~35~K   &  1.27 & 2.8   &  6.7 \\
~70~K   &  1.71 & 0.41  &  8.5 \\
120~K  &  2.22 & 0.275 & 10.0 \\
\end{tabular}
\end{ruledtabular}
\end{table}

\begin{figure}[ht]
\includegraphics[scale=0.5]{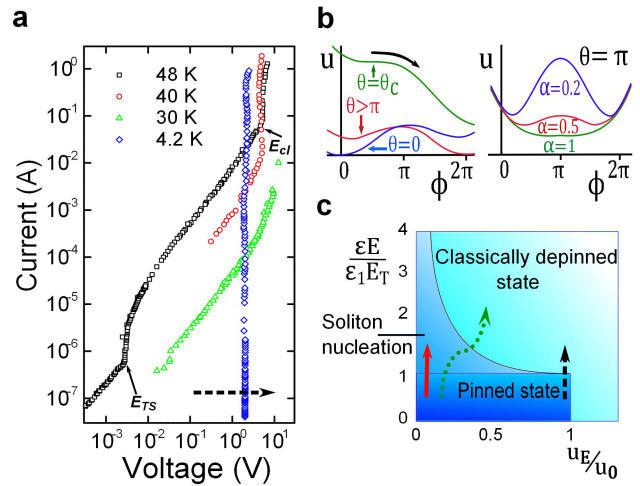}
\caption{(color online)~\textbf{a}.~Blue bronze $I-V$ curves~\cite{ Mihaly1987911}.
\textbf{b~(left)}:~$u$ vs. $\phi$, showing  $\theta\geq\pi$  quantum instability and  $\theta\geq\theta_c$  classical depinning.
\textbf{b~(right)}:~$u(\phi)$ at $\theta=\pi $ for several~$\alpha=u_E/u_0$. 
\textbf{c}.~Phase diagram showing pinned, soliton nucleation, \&~classically depinned states. Solid arrow:~$u_E/u_0<<1$  for which soliton pair creation dominates. Dotted arrow shows both soliton nucleation and classical depinning.
Since $u_E\propto 1/\epsilon$, the path curves to the left (right) if~$\epsilon$ increases (decreases). Dashed arrows (\textbf{a} and \textbf{c}):~classical depinning dominates.\label{fig3}}
\end{figure}

Some CDW crystals exhibit more than one threshold within certain temperature ranges~\cite{Mihaly1987911,1990JPCM....2.8327I} (Fig.~\ref{fig3}a). The two major thresholds emerge naturally, provided the nucleated soliton conductance is sufficiently small for~$\theta$ to be treated quasi-statically, i.e.~$\theta=\pi\epsilon E/\epsilon_1E_T$, where~$\epsilon_1=\epsilon (E\approx E_T)$. We interpret the low- and high-field thresholds as due to soliton nucleation and classical depinning, respectively. Fig.~\ref{fig3}b~(left) illustrates the quantum $(\theta \geq \pi )$ and classical~$(\theta \geq \theta_c)$ instabilities, where $\theta_c(\alpha)\cong\alpha^{-1}+\pi/2$    when $\alpha=u_E/u_0 <<1$. Fig.~\ref{fig3}b~(right) plots $u(\phi)$ when~$\theta=\pi$ for several values of~$\alpha$. Figure~\ref{fig3}c shows the resulting  $\theta$  vs. $u_E/u_0$  phase diagram, which illustrates the pinned state, for  $\theta<\pi$  and $u_E/u_0<1$, a region $(\pi\leq\theta\leq\theta_c )$ in which soliton nucleation occurs, and a high field classical depinning region~$(\theta\geq\theta_c)$.

The observed flat ac responses~\cite{PhysRevB.31.5229} and small phase displacements ~\cite{PhysRevLett.56.663} below threshold in NbSe$_3$ and TaS$_3$ suggest $u_E/u_0~<<~1$  (solid arrow, Fig.~\ref{fig3}c), where soliton nucleation dominates. For example, computed~\cite{PhysRevLett.84.1555} phase displacements $\left\langle \phi\right\rangle$ below threshold compare favorably to the measured 2$^{\circ}$ phase displacement obtained from NMR experiments~\cite{PhysRevLett.56.663} on NbSe$_3$, provided $u_E/u_0$ is taken to be 0.015~\cite{PhysRevLett.84.1555}. Using, $u_E/u_0= 2\pi E_T/E_{cl}$, the~48~K blue bronze data~\cite{Mihaly1987911} in  Fig.~\ref{fig3}a  suggests a similar value of about~0.01. However, the soliton nucleation threshold field $E_T$, and consequently $u_E/u_0$, increase with decreasing temperature, whereas the classical threshold $E_{cl}$ shows a weak temperature dependence. We interpret the change in soliton nucleation threshold (which scales inversely with $\epsilon$) as due to a reduction in ~$\epsilon$ as the normal carrier concentration goes down with decreasing temperature. At 4~K the normal carriers become frozen out, resulting in a relatively low~$\epsilon$ and sufficiently high~$u_E/u_0$ and $E_T$ for classical depinning to dominate (dashed arrows in Figs.~\ref{fig3}a and~\ref{fig3}c).

A potential topic of interest is to study coupling of static and dynamic vector potentials to the relative phase $\delta_1-\delta_0$ between order parameters. This will be critical, both for understanding the CDW ring Aharonov-Bohm experiments~\cite{Tsubota2009416} and for interpreting ac response experiments~\cite{PhysRevB.31.5229}, which show remarkable agreement with photon-assisted tunneling theory. Another variation of the model represents multiple DW domains, due to random pinning, as a network of many resistively shunted junctions of type shown in  Fig.~\ref{fig1}c.

Density wave transport is one of very few known cases of correlated transport of macroscopic numbers of electrons over long distances--- the only known example of large-scale collective electron transport at biological temperatures (e.g. NbS$_3$, with~$T_{Peierls} \sim$~360~K~\cite{2009ApPhL..94o2112Z}). It is hoped that this Letter will revitalize this important branch of condensed matter physics, for which quantum principles have largely been ignored by most for the past thirty years. Additional areas of impact include improved understanding of other correlated electron systems, flux vortex nucleation, tunneling in quantum cosmology~\cite{springerlink:10.1007/BF02790571}, and $\theta = \pi$ instabilities in spontaneous CP violation~\cite{PhysRevD.61.114009}. Finally, understanding of the quantum behavior of solitons could potentially lead to topologically robust forms of quantum information processing.

\begin{acknowledgments}
The authors acknowledge technical assistance by Rabi Ebrahim and Jarek Wosik.
JHM and AIW acknowledge support by R21CA133153 from NIH (NCI) and by ARRA supplement: 3R21 CA133153-03S1 (NIH, NCI).
AMG and ZT acknowledge support by NSF (CHE-0616805) and the R.A.~Welch Foundation (E-1297).
Additional support was provided by the State of Texas through the Texas Center for Superconductivity at the University of Houston.
\end{acknowledgments}


\end{document}


\title{Supplemental Material for:\\Correlated Quantum Transport of Density Wave Electrons}

\author{J.~H.~Miller,~Jr.}
\email[]{jhmiller@uh.edu}
\affiliation{Department of Physics, University of Houston, Houston, Texas 77204-5005 USA}
\affiliation{Texas Center for Superconductivity, University of Houston, Houston, Texas 77204-5002 USA}

\author{A.~I.~Wijesinghe}
\affiliation{Department of Physics, University of Houston, Houston, Texas 77204-5005 USA}
\affiliation{Texas Center for Superconductivity, University of Houston, Houston, Texas 77204-5002 USA}
\author{Z. Tang}
\affiliation{Department of Chemistry, University of Houston, Houston, Texas 77204-5003 USA}
\author{A.~M.~Guloy}
\affiliation{Texas Center for Superconductivity, University of Houston, Houston, Texas 77204-5002 USA}
\affiliation{Department of Chemistry, University of Houston, Houston, Texas 77204-5003 USA}

\date{\today}

\maketitle


\textit{\textbf{Dislocation pair nucleation}}. Density waves are often highly anisotropic, where the dielectric response, $\epsilon_{||}$, along the chain direction is much greater than that, $\epsilon_{\bot}$, in the perpendicular directions. In this case, the Coulomb blockade threshold for nucleation of a $2\pi $ soliton-antisoliton (dislocation-antidislocation) pair (Fig.~S-1a) is comparable to that for domain wall pair creation, within a factor of $\sim 1/2$. This is seen by starting with the Maxwell equation: $\nabla\cdot \textbf{D}=\rho$, where (using the summation convention): $D_i=\epsilon_0\epsilon_{ij}E_j$. Here $\epsilon_{ij}$ is the relative dielectric tensor, which is diagonal with elements $\epsilon_{||}\equiv\epsilon_{xx}$,  $\epsilon_{\bot y}\equiv\epsilon_{yy}$ and   $\epsilon_{\bot z}\equiv\epsilon_{zz}$ if the axes $i, j = x, y,$ and $z$ are along principal crystallographic directions. Introducing rescaled variables,$x'=x/\epsilon_{||}$, $y'=y/\epsilon_{\bot y}$, and $z'=z/\epsilon_{\bot z}$, yields: $\nabla'\cdot \textbf{E}=\rho/\epsilon_0$, resulting in the modified charge distribution, electrostatic potential, and field lines shown in Fig. S-1b, which assumes that $\epsilon_{||}>>\epsilon_{\bot x,y}$. In the rescaled coordinate system of Fig.~S-1b, the dislocation pair looks like a parallel plate capacitor that produces an internal field $E^*=2e/2\epsilon A_{ch}$, where $A_{ch}$ is the cross-sectional area of a DW chain and $\epsilon=\epsilon_{||}\epsilon_{0}$. Fig.~S-1c illustrates the aggregation of many $2\pi$ dislocations of charge $2e$ into soliton domain walls, each of which may behave as a quantum fluid and between which the bubble of `true vacuum' grows.

\textit{\textbf{Impurity and temperature dependences of quantum and classical threshold fields}}. The Coulomb blockade threshold of a density wave scales inversely with low frequency dielectric response $\epsilon$ which, in turn, is strongly influenced by pinning due to impurity concentration $n_i$, normal carrier concentration, and temperature $T$. The Gr\"{u}ner relation between dielectric response and threshold, for the simplest classical model, is given by~\cite{gruner1998,gruner1994}: $\epsilon E_{cl} = 4\pi en_{ch}\rho _c$, where $n_{ch} = N/A$ and $E_{cl}$ is the classical depinning field. 
The Coulomb blockade threshold $E_T = Q_0/\epsilon A$ leads to a similar relationship: $\epsilon E_T = en_{ch}\rho _c$, with the result, $E_T = E_{cl}/4\pi$,   which scales as $n_i^2$ for a density wave weakly pinned by impurities. Screening by normal carriers, both thermally activated and due to incomplete Peierls gaps in materials such as NbSe$_3$, further enhance $\epsilon$ and reduce the ratio $E_T/E_{cl}$. For a fixed $n_i$, the temperature dependence of $\epsilon(T)$ leads (inversely) to the observed temperature dependence of $E_T$. This leads to rather complex temperature dependence in materials such as NbSe$_3$, for example, which undergoes two separate Peierls transitions.

The possibility of additional $I-V$ inflection points at intermediate fields in the soliton nucleation region below the classical depinning field is suggested by some experiments on TaS$_3$~\cite{prl781098}, and might occur if the first nucleated solitons were impeded from rapidly reaching the contacts, perhaps by disorder.  A second bubble bounded by solitons could then nucleate within the first bubble.  The additional electrostatic energy would be: 
$\frac{1}{2}\epsilon A \ell [(E-2E^{*'})^2- (E-E^{*'})^2] = \epsilon A \ell E^{*'}[\frac{3}{2}E^{*'}-E]^2$, which is positive unless $E$ exceeds a higher Coulomb blockade threshold, $E_{T3}= \frac{3}{2}E^{*'} = 3\frac{\epsilon_1}{\epsilon_3} E_{T1}$.   (Note that, in general, $\epsilon _3~\equiv~\epsilon (E = E_{T3})\neq\epsilon_1\epsilon (E=E_{T1})$.) In a similar fashion, a series of higher Coulomb blockade thresholds $E_{Tn}$ could  exist, provided disorder or other mechanisms prevented solitons from annihilating or reaching the contacts quickly. Furthermore, 4$\pi$, 6$\pi$, etc. soliton domain walls might nucleate, presumably with much lower probability, causing additional inflection points at $\theta=2\pi n$. A generalization of the Gr\"{u}ner relation would then have the form: $\epsilon_n E_{Tn}=nen_{ch}\rho_c$, where $n$ is an integer. Such possibilities can potentially be explored in crystals with both weak and strong pinning centers, perhaps through ion irradiation of initially weakly pinned samples.

\textit{\textbf{Discussion of relevant energy scales}}. The microscopic quantum soliton energy gap (per dressed electron pair) $\Delta_{\phi}$ can be estimated from the measured Zener field $E_0$, typically $\sim$~10~V/m. Using a phason velocity $v_0=(m/M_F)^{1/2}v_F$~$\sim3\times10^5$~m/s, where $M_F/m$ is the Fr\"{o}lich mass ratio~\cite{gruner1998,gruner1994}, and $E_0\sim\Delta_{\phi}^2/\hbar v_0e$, yields $\Delta_{\phi}\sim4\times10^{-5}$~eV. Although this microscopic energy is extremely small, $\Delta_{\phi}\ll~kT$, both the initial and emerging macrostates become macroscopically occupied with substantial condensation energies due the large number ($\sim$~10$^9$) of interacting parallel chains. Thus the condensed quantum solitons and antisolitons in the emerging macrostate are effectively trapped in quantum fluids, preventing thermal excitations except across the Peierls gap, $\Delta_P>kT$, typically several 10's of meV. 

Two analogies are helpful here. One is coherent Josephson coupling between two superconducting order parameters that are superimposed in space, as discussed by Leggett~\cite{Leggett}. In this case, it is immaterial whether the macrostates are separated by energies per electron pair that are small compared to~$kT$, since thermal excitations only occur across either of the two BCS energy gaps. Another analogy is that of quantum creep of superfluid helium out of a container. If the height of the rim is, for example, $d\sim$1~cm above the liquid surface inside the container, then the gravitational energy barrier for each helium atom, $mgd$ is only $\sim4\times10^{-9}$~eV, which is small compared to $kT$ even for temperatures as low as 1~mK. Nevertheless, the superfluid helium atoms remain trapped in the quantum fluid and prevented by the condensation energy from thermally hopping over the barrier as individual particles.

\renewcommand{\figurename}{FIG S-}
\begin{figure*}[ht]
\includegraphics[scale=.8]{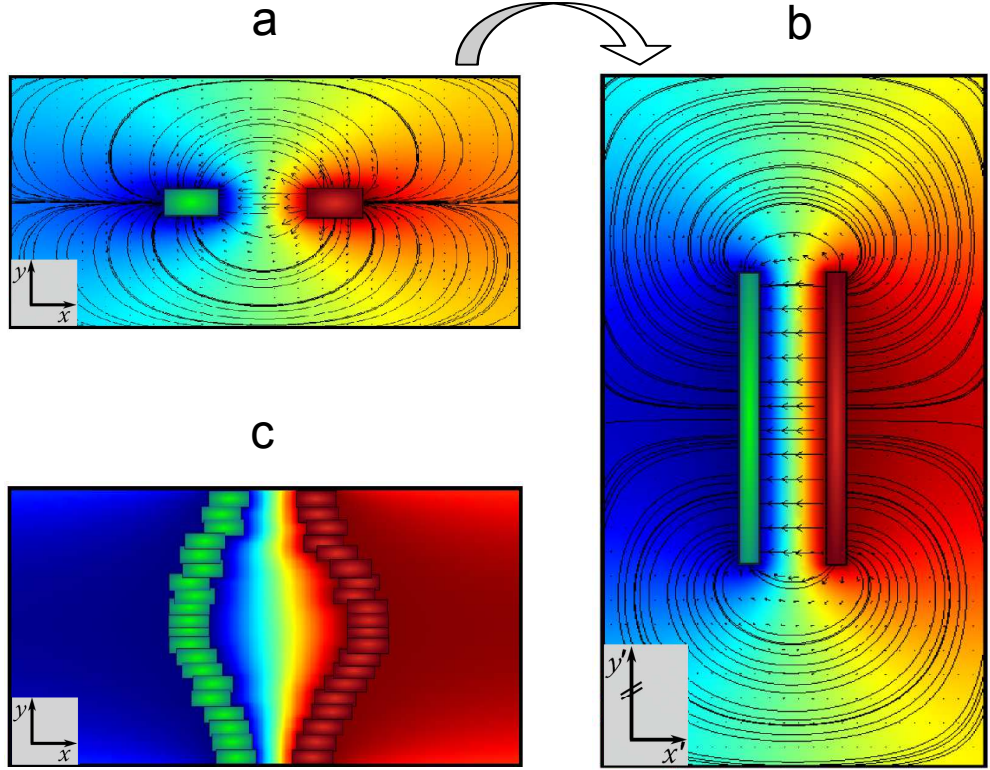}
\caption{\textbf{(a)}.~COMSOL Multiphysics (COMSOL,~Inc.) simulation of the electrostatic potential (red =~positive, blue =~negative) and electric field lines for an electric dipole consisting of dislocations represented as positive and negative rectangular charge distributions.
\textbf{(b)}.~COMSOL simulation for the same system, but with highly anisotropic dielectric constants, in the rescaled coordinate system.  The apparent distance between the charges is greatly reduced along the $x'$ direction, yielding an arrangement resembling a parallel plate capacitor. 
\textbf{(c)}.~In the original coordinate system, aggregation of many dislocations into `domain walls', between which the bubble of `true vacuum' grows as the domain walls are driven toward the contacts by the externally applied electric field.\label{figs1}}
\end{figure*}